\documentclass[12pt,a4paper]{article}
\usepackage{latexsym}
\usepackage{enumerate}
\usepackage{calc}
\usepackage{amsmath,amsthm}
\usepackage[mathscr]{euscript}
\usepackage{amssymb}
\usepackage[dvips]{graphicx}
\theoremstyle{plain}
\newtheorem{twr}{Theorem}[section]
\newtheorem{lemat}[twr]{Lemma}

\theoremstyle{definition}
\newtheorem{defin}[twr]{Definition}

\theoremstyle{remark}

\swapnumbers 

\newcommand{\Var}{\operatorname{Var}}

\newcommand{\BIB}[7]{\bibitem{#1} #2, \emph{#3}, #4 {\bf #5} (#6), #7.}

\newcommand{\BIb}[5]{\bibitem{#1} #2, \emph{#3}, #4, #5.}

\begin{document}

\author{Edward W. Piotrowski\\ Institute of Mathematics,
University of Bia\l ystok,\\ Lipowa 41, Pl 15424 Bia\l ystok,
Poland\\ e-mail: ep@alpha.uwb.edu.pl\\
 Ma\l gorzata Schroeder\\ Institute of Mathematics, University of Bia\l ystok, \\ Akademicka 2, Pl 15267 Bia\l ystok, Poland\\ e-mail: mszuli@math.uwb.edu.pl\\
 Anna Szczypi\'nska\\ Institute of Physics, University of Silesia, \\ Uniwersytecka
4, Pl 40007 Katowice, Poland \\ e-mail:
zanna12@wp.pl}
\title{The price of bond and European option on bond without credit risk.\\ Classical look and its quantum extension.}
\date{} 
\maketitle

\begin{abstract}
${}$\indent In this paper we compare two classical one-factor diffusion models
which are used to model the term structure of interest rates.
One of them is based on the Wiener-Bachelier process while the second one is based on the Ornstein-Uhlenbeck process. We show essential differences between the prices of European call options on a zero-coupon bond in these models.
\end{abstract}
\vspace{5mm}

\newpage
\begin{flushleft}
{\Large\bf Introduction}
\end{flushleft}

The bond market forms a very important segment of financial markets. However, modelling of it is difficult because the bond prices depend on both the present situation of the market
and the time to maturity. The prices of bonds are expressed in terms of various interest rates and yields, so understanding bonds pricing is equivalent to understanding interest rate
behaviour. The term structure of interest rates is defined as the dependence between interest rates and maturities. 
We propose two models:\par
1) the Merton model based on the Wiener-Bachelier process,\par
2) the Vasicek model based on the quantum Ornstein-Uhlenbeck process.\footnote{In the quantum
game theory Ornstein-Uhlenbeck process has the interpretation of non-unitary tactics resulting in a new strategy. These strategies are the Hilbert's spaces vectors. Quantum strategies create unique opportunities for making profits during intervals shorter than the
characteristic thresholds for the Brown particle, see \cite {6a,7}.}\\
In the above models we find the formula for the zero-coupon bond price and we price 
the option on this bond. In the first section we propose general zero-coupon bond pricing model
and we give some information about forward contracts with the help of which we price options on zero-coupon bonds. Then, we discuss the Merton model while in the third one we discuss the Vasicek model. Finally, we compare both models.\\

We assume that the market is effective without an opportunity of arbitrage and the face value of any bond equals one. Let $B_t^T$ be the price of the zero-coupon bond at some date $t$ with the time to maturity $T\geq t$. Let $y_t^T$ denotes the discount rate called {\it zero-coupon rate} or {\it spot rate} for date $T$ in continuous capitalisation \cite{10}. Then $B_t^T$ is related to the continuously compounded zero-coupon rate by 
\begin{equation}
B_t^T = e^{-(T-t)y_t^T},
\end{equation}
hence
\begin{equation}
y_t^T = -\frac{1}{T-t} \ln B_t^T.
\end{equation}
We call the function $T \longrightarrow B^T_t$ {\it the discount function} and
the transformation $T \longrightarrow y_t^T$ {\it the yield curve}.\\

{\it The short-term interest rate} is given by the formula $$r_t =
\lim_{\tau \to 0+} y_t^{t+\tau}.$$
We introduce few notions which are connected with the derivatives. First, we consider options on the zero-coupon bonds. Let us consider European options. We must clearly distinguish between the call option and the put option. The time of maturity of the option is denoted by $T$. Let $C_t$ denotes the time $t$ price of an European call
option on the zero-coupon bond, which gives a payment of $1$ at time $S$, where $S \geq T$, with the exercise price of $K$. Similarly, let $\pi_t$ denotes the price of the corresponding put option. The prices of these options at maturity are equal to $C_T = \max(B_T^S - K,0)$ and $\pi_T = \max(K - B_T^S,0)$\footnote{ $B_t^T$ always
denotes the price of the zero-coupon bond at the time $t$ with the time to maturity $T$.}.
The prices of European call and put option on the zero-coupon bonds fulfil the {\it put-call parity} relation \cite{1}
\begin{equation}\label{parytet kupna sprzeda¿y}
C_t + K B_t^T = \pi_t + B_t^S.
\end{equation}
As the result, we can restrict ourself to pricing of European call options.\par
Another derivatives are {\it forward contracts} on the zero-coupon bonds. 
If the underlying variable is the price of the zero-coupon bond, the essential part by the pricing of the forward contracts plays the present prices of the zero-coupon bonds only.
If there is no arbitrage, the time $t$ value $V_t^{T,S}$ of the forward contract on the zero-coupon bond, with the delivery date $T$ and the delivery price $K$, is given by the formula  \cite{1}
\begin{equation}\label{for1254}
V_t^{T,S} = B_t^S - K B_t^T.
\end{equation}
$S > T$ is the maturity date of the underlying bond. For forwards contracted upon at time $t$, the delivery price $K$ is set so that the value of the forward at time $t$ is zero ($V_t^{T,S}=0$). This value of $K$ is called the {\it forward 
price} $F_t^{T,S}$. Solving (\ref{for1254}) we obtain that 
\begin{equation}\label{forwards}
F_t^{T,S} = \frac{B_t^S}{B_t^T}.
\end{equation}

\section{The pricing in affine models}

${}$\indent We assume that for considered market model, there is a risk-neutral probability measure (or equivalent martingale measure) $P^*$ and  
a one-dimensional Wiener process $W^*$ under measure $P^*$. If the risk-neutral probability measure exists, then there is also the {\it $T$-forward martingale measure} $P^T$ and then $W^T$ is the Wiener process under measure $P^T$. If there exists a spot martingale measure, then the market is arbitrage free. Part of the market characteristic fulfils the short rate under the risk-neutral probability  measure (i.e. the spot martingale measure). The model in which this rate is represented by the process $(r_t)_{t>0}$, that fulfils the stochastic equation
\begin{equation}\label{def:rów stoch}
dr_t = \mu(r_t,t)dt + \sigma(r_t,t) dW^*_t,
\end{equation}
where $\mu:\mathbb{R}\times[0,T]\longrightarrow\mathbb{R}$ and $\sigma:\mathbb{R}\times[0,T]\longrightarrow\mathbb{R}^d$, we call the \textit{affine model}. Let us consider that $\mu_t=\alpha_1-\alpha_2\cdot r_t$ and $\sigma_t=\sqrt{\beta_1+\beta_2\cdot r_t}$, where $\alpha_1,\alpha_2,\beta_1,\beta_2$ are constant. We know that the discount price of the zero-coupon bond is given by the formula \cite{2}
\begin{equation}
B^T_t = E^{P^*}\left[e^{-\int_t^T r_s ds}|\mathcal F_t\right].
\end{equation}
With the help of this assumption, we can calculate the formula for the discount price of the zero-coupon bond, see \cite {3}. This price can be written as a function of time and the current short rate $r$. 
\begin{twr}${}$\\
In an affine model, in which the short rate $r$ is described by (\ref{def:rów stoch}),
the time $t$ price of the zero-coupon with the time to maturity $T$ and face value $1$ is equal to 
\begin{equation}\label{cena obligacji}
B^T_t = e^{-a(T-t) - r b(T-t)}.
\end{equation}
The functions $a,b: [0,T] \longrightarrow \mathbb{R}$ fulfil the following system of ordinary differential equations:
\begin{align*}
\frac12 \beta_2 b(t)^2 + \alpha_2 b(t) + b'(t) - 1 &= 0,\\
a'(t)-\alpha_1 b(t) + \frac12 \beta_1 b(t)^2 &= 0,
\end{align*}
for $t \in (0,T)$ and $a(0) = b(0) = 0$. \label{2.1co}
\end{twr}
Proof.\\  The proof is based on the Feynman-Kac theorem, see (\ref{twierdzenie feymana kaca}) in Appendix. From theorem (\ref{twierdzenie feymana kaca}) follows that $B^T_t$, given by the formula (\ref{cena obligacji}), fulfils the partial differential equation
$$
\frac{\partial B^T_t}{\partial t} + (\alpha_1 - \alpha_2 r)
\frac{\partial B^T_t}{\partial r} +\frac12 (\beta_1 + \beta_2
r) \frac{\partial^2 B^T_t}{\partial r^2} - r B^T_t = 0.
$$
The relevant derivatives are
\begin{eqnarray*}
\frac{\partial B^T_t}{\partial t} &= & B^T_t (a'(T-t) + r b'(T-t)),\\
\frac{\partial B^T_t}{\partial r} &= & - B^T_t b(T-t),\\
\frac{\partial^2 B^T_t}{\partial r^2} &= & B^T_t b(T-t)^2.
\end{eqnarray*}
After substituting these formulas into (\ref{twierdzenie feymana kaca}), we obtain
$$
a'(T-t) + r b'(T-t) - (\alpha_1 - \alpha_2 r) b(T-t) + \frac12
b(T-t)^2 (\beta_1 + \beta_2 r)^2 = 0.
$$
After gathering terms involving $r$, we obtain
\begin{multline*}
(a'(T-t) - \alpha_1 b(T-t) + \frac12 \beta_1 b(T-t)^2) +\\
+ (\frac12 \beta_2 b(T-t)^2 + \alpha_2 b(T-t) + b'(T-t) - 1)r = 0.
\end{multline*}
The last equation must be fulfilled for arbitrary $r$, so the expressions in the brackets must be identically $0$. Hence, we find equations which the function $a$ and $b$ must fulfil.
Because $B^T_T \equiv 1$, so $a(0) + b(0) r \equiv 0$ and hence $a(0) = b(0) = 0$. 
We see that the function (\ref{cena obligacji}) fulfils the Feynman-Kac equation.
\begin{flushright}
 $\square$\\
\end{flushright}
Let us observe that, if we have the function $b$ from the above formula, 
we can calculate the function $a$ from the formula\footnote{If $a'(t) = \alpha_1 b(t) - \frac12 \beta_1
b(t)^2$, so $a(t) = a(0) + \alpha_1 \int_0^t b(s)ds - \frac12
\beta_1 \int_0^t b(s)^2ds $. Let us take into consideration that $a(0) = 0$.}
\begin{equation}\label{wzór na a}
a(t) = \alpha_1 \int_0^t b(s)ds - \frac12 \beta_1 \int_0^t
b(s)^2ds.
\end{equation}
If we want to determine the price of the bond, we have to find the solution
of the differential equation given by the formula
$$\frac12 \beta_2 b(t)^2 + \alpha_2 b(t) + b'(t) - 1 = 0, \qquad b(0)= 0.$$
Now, let us observe that $b(t) > 0$ for $t > 0$.
Above all, from the continuity of $b$, the conditions $b(0)=0$ and $b'(t) = 1
- \frac12 \beta_2 b(t)^2 - \alpha_2 b(t)$ follow that $b'(t) > 0$ in
the vicinity of $0$. From this, $b$ grows up in the vicinity of $0$, that is
$b(t)> 0$ for small $t > 0$. As if $b(t) \leq 0$ for
certain $t> 0$, there is a point $t_1 > 0$ from the {\it Intermediate Value Theorem} \cite{4}, such that $b(t_1) = 0$. From the {\it Rolle theorem} \cite{4}, we obtain $b(0) =
b(t_1) = 0$, so there must exist $t_0 > 0$ such that $b'(t_0) = 0$. If we substitute
$t_0$ to the differential equation fulfilled by $b$,
we obtain the equality $\frac12 \beta_2 y_0^2 + \alpha_2 y_0 - 1 = 0$
for $y_0 = b(t_0)$. Let us consider constant function $b_0(t) =
y_0$. It fulfils the same differential equation, so
$b$. Furthermore, $b(t_0) = b_0(t_0)$, that means that both functions fulfil
the same initial-value problem. From the uniqueness of solution we obtain that
$b \equiv b_0$ and $b_0(0) = b(0)$, that is, $y_0 = 0$, which is
impossible because $0$ does not fulfil the quadratic equation which is satisfied by $y_0$. We obtain contradiction, so $b(t)> 0$ for $t > 0$. It means that the function $r\longrightarrow B^T_t$ is strictly decreasing.\par
In the further parts of our paper we will use the formula for the volatility of the process
$B^T$ under {\it $T$-forward martingale measure}. It looks the same as for measure
$P^*$. From the Feynman-Kac theorem, it is given by the formula
$$
\sigma^T_t = \frac{\frac{\partial B^T_t}{\partial r}}{B^T_t}
\beta_t
$$
and in our specific situation $\beta_t  =
\sqrt{\beta_1 + \beta_2 r}$. Using the formula (\ref{cena obligacji}), we obtain
\begin{equation}
\sigma^T_t = \frac{-b(T-t) B^T_t}{B^T_t} \beta_t =
-b(T-t) \beta_t.
\end{equation}%
Let $\mu^T$ denotes the relative drift of the process $B^T$ under measure $P^T$.

\section{The Merton model}

${}$\indent The first dynamic, affine and continuous time model of the term structure of the interest rate was described by Merton \cite {5}. The short rate follows a generalised Brownian motion under the spot martingale measure: 
$$
dr_t = \varphi dt + \sigma d{W^*_t},
$$
where $\sigma $ and $\varphi$ are constant. We begin by finding the price of the zero-coupon bond and then we calculate the price of the option. We see that the price of the bond in this model is given by the formula  (see (\ref {cena obligacji}))
$$
B^T_t = e^{-a(T-t) - r b(T-t)},
$$
where the function $b$ fulfils the simple ordinary differential equation 
$b'(t) = 1$ with $b(0) = 0.$
So, $b(t) = t$ and the function $a$ is given by the formula (see (\ref{wzór na a}))
$$
a(t) = \varphi \int_0^t s ds - \frac12 \sigma^2 \int_0^t s^2 ds =
\frac12 \varphi t^2 - \frac16 \sigma^2 t^3,
$$
therefore
\begin{equation}\label{wzór na obligacjê Merton}
 B^T_t = e^{-\frac12 \varphi (T-t)^2 +\frac16 \sigma^2
(T-t)^3 - (T-t)r}.
\end{equation}
In order to price the option, we use the theorem (\ref{opcje}), see Appendix.
The price of an European call option on the zero-coupon bond with the expiration date $S$, the exercise price $K$, and maturing at the time $T$ is given by, see (\ref{opcje!}) in Appendix,
\begin{equation}\label{cena opcji}
C_t = B^T_t \left(E^{P^T}\left[\max(B^S_T -
K,0)|\mathcal F_t \right]\right)(r).
\end{equation}
From (\ref{forwards}) we know that the \emph{forward price} fulfils the formula:
$$
F^{T,S}_t = \frac{B^S_t}{B^T_t},
$$
hence, in particular $F^{T,S}_T = B^S_T$. The \emph{forward price} 
$(F^{T,S}_t)_{t\geq 0}$ is a martingale under the $T$-\emph{forward martingale measure}, what means that the drift rate equals $0$ under this probability measure. Simultaneously, this process is a quotient of the prices of two zero coupon bonds. In the diffusion equations for $B^T$ and $B^S$, the relative volatilities of the bond are equal to $\sigma^T_t =
-\sigma \cdot (T-t)$ and $\sigma^S_t = -\sigma \cdot (S-t)$ respectively, so that
\begin{align*}
\textrm{d}B^T_t &= \mu^T_t B^T_t dt
-\sigma (T-t) B^T_t d{W^T_t},\\
\textrm{d}B^S_t &= \mu^S_t B^S_t dt -\sigma (S-t) B^S_t
d{W^T_t}.
\end{align*}
By an application of the It\^o lemma, see Appendix (\ref{wzór Ito}), for functions of multiple stochastic processes and  because of that, the drift of the process $F^{T,S}$ equals $0$ under the $T$-\emph{forward martingale measure}, we obtain
\begin{align*}
\textrm{d}F^{T,S}_t &= \left[\frac{\partial{F^{T,S}_t}}{\partial
B^T_t} \sigma^T_t B^T_t +
\frac{\partial{F^{T,S}_t}}{\partial B^S_t}
\sigma^S_t B^S_t\right]
d{W^T_t}\\
&= \left[-\frac{B^S_t}{(B^T_t)^2} \sigma^T_t B^T_t
+ \frac{1}{B^T_t} \sigma^S_t B^S_t\right] d{W^T_t}\\
&= \left(\sigma^S_t - \sigma^T_t\right) F^{T,S}_t d{W^T_t}
= -\sigma (S-T) F^{T,S}_t d{W^T_t}.
\end{align*}
This means that, $F^{T,S}$ follows a geometric Brownian motion, see Appendix (\ref{def:geom Brown}), with $\sigma^{T,S}(t) = -\sigma \cdot (S - T)$. Hence, the random variable 
$X := \ln \frac{F^{S,T}_T}{F^{S,T}_t}$ is normally distributed and
\begin{equation}\label{wzór na x}
X = - \frac12 \int_t^T \sigma^{T,S}(u)^2 du + \int_t^T
\sigma^{T,S}(u) d{W_u^T},
\end{equation}
what result from (\ref{twr:geom Brown}), see Appendix. Moreover, 
\begin{eqnarray*}
E(X) = -\frac12 \int_t^T \sigma^{T,S}(u)^2 du =
-\frac12 \sigma^2 (S - T)^2 (T - t),\\
v(t,T,S) := \sqrt{Var(X)} = \sqrt{\int_t^T \sigma^{T,S}(u)^2 du} =
\sigma (S - T) \sqrt{T - t}.
\end{eqnarray*}
Let us consider the function $g: \mathbb{R} \times \mathbb{R_+}
\ni (x,y) \longrightarrow \max(e^x y - K,0)$ and a random variable $Z = e^X$.
From the formula (\ref{cena opcji}) and with the help of that $B^S_T =F^{T,S}_T$, we obtain
\begin{eqnarray*}
C_t &= B^T_t\left(E^{P^T}\left[\max(F^{T,S}_T - K,0)|\mathcal F_t\right]\right)(r) =\\
&= B^T_t \left(E^{P^T}\left[g(X,F^{T,S}_t)|\mathcal
F_t\right]\right)(r).
\end{eqnarray*}
From the equation (\ref{wzór na x}) we have that $X$ is independent of $\mathcal F_t$ and
$F^{T,S}_t$ is $\mathcal F_t$-\emph {measurable}, because the process $F^{T,S}$ is a martingale. Hence, we can use the lemma (\ref{borelowska})\footnote{Let us observe that $g(X,F^{T,S}_t) = \max(F^{T,S}_T -
K,0)$, so that $g(X,F^{T,S})$ is bounded, because $0 \leq F^{T,S}
\leq 1$.}, see Appendix. By the using the lemma (\ref{lognorm}), see Appendix, we can calculate the expected value [$y > 0$ !]
\begin{multline*}
E(g(X,y)) = E(\max(Z \cdot y - K,0))
= y E\left(\max\left(Z - \frac{K}{y},0\right)\right) =\\
= y\left[e^{E(X) + \frac12 v(t,T,S)^2} N\left(\frac{E(X) -
\ln\frac{K}{y}}{v(t,T,S)} + v(t,T,S)\right) - \frac{K}{y}
N\left(\frac{E(X) - \ln\frac{K}{y}}{v(t,T,S)}\right)\right] =\\
= y e^0 N\left(-\frac12 v(t,T,S) +\frac{1}{v(t,T,S)}
\ln\frac{y}{K} + v(t,T,S)\right)+\\
- K N\left(-\frac12 v(t,T,S)
+ \frac{1}{v(t,T,S)}\ln\frac{y}{K}\right) =\\
= y N\left(\frac{1}{v(t,T,S)} \ln\frac{y}{K} + \frac12
v(t,T,S)\right) - K N\left(\frac{1}{v(t,T,S)} \ln\frac{y}{K} -
\frac12 v(t,T,S)\right).
\end{multline*}
If we substitute to the above equation $y = F^{T,S}_t$, we obtain
\begin{multline*}
\left(E^{P^T}\left[g(X,F^{T,S}_t)|\mathcal F_t\right]\right)(r) =\\
= F^{T,S}_t N\left(\frac{1}{v(t,T,S)} \ln\frac{F^{T,S}_t}{K}
+ \frac12 v(t,T,S)\right) +\\ - K N\left(\frac{1}{v(t,T,S)}
\ln\frac{F^{T,S}_t}{K} - \frac12 v(t,T,S)\right)
\end{multline*}
and  finally 
\begin{align*}
C_t &= B^T_t\left[F^{T,S}_t N(d_1) - K N(d_2)\right]= B^S_t N(d_1) - K B^T_t N(d_2),
\end{align*}
where
\begin{align*}
d_1 &= \frac{1}{v(t,T,S)} \ln\left(\frac{B^S_t}{K B^T_t}\right) + \frac12 v(t,T,S),\\
d_2 &= \frac{1}{v(t,T,S)} \ln\left(\frac{B^S_t}{K B^T_t}\right) - \frac12 v(t,T,S),\\
v(t,T,S) &= \sigma (S - T) \sqrt{T - t}.
\end{align*}

\section{The Vasicek Model}

\indent In the paper \cite{6} Vasicek proposed a mean-reverting version of the Ornstein-Uhlenbeck process for the short term rate. Specifically, under the risk-neutral measure $P^*$, $r_t$ is given by
$$
\textrm{d}r_t = \kappa (\theta - r_t)dt + \sigma d{W^*_t},
$$
where $\kappa,\ \theta$ and $\sigma$ are positive constants. The parameter $\theta$
denotes \emph{long-term level of the short rate}, because the rate $r_t$ pull towards a long-term level of $\theta$, $\kappa$ determines the speed of adjustment, and $\sigma$ is the average deviation of the rate of return.\par
Similarly to the Merton model, at the first we calculate the formula for the zero-coupon bond price. From the theorem \ref{2.1co} we know that
$$
B^T_t = e^{-a(T-t) - r b(T-t)},$$
but in the described model we have
$$\kappa b(t) + b'(t) - 1 = 0, \qquad b(0) = 0\,.$$
Hence,
$$
b(t) = \frac{1}{\kappa}(1 - e^{-\kappa t}).
$$
From (\ref{wzór na a}) we obtain that
\begin{multline*}
a(t) = \kappa \theta \int_0^t b(s)ds - \frac12 \sigma^2 \int_0^t
b(s)^2 ds = \\ = \theta \int_0^t (1 - e^{-\kappa s}) ds -
\frac{\sigma^2}{2 \kappa^2} \int_0^t (1 -2 e^{-\kappa t}
+ e^{-2\kappa t})ds =\\
= (\theta - \frac{\sigma^2}{2 \kappa^2})(t - b(t)) +
\frac{\sigma^2}{4 \kappa} b(t)^2.
\end{multline*}
If we substitute $a(t)$ and $b(t)$ to the formula (\ref{cena obligacji}), we obtain analogous formula for the bond price to the Merton model, see (\ref{wzór na obligacjê Merton}).\\

We calculate the formula for the call option price similar to the Merton model.
Let us use the basic formula
$$
C_t = B^T_t E^{P^T}\left[\max(B^S_T -
K,0)|\mathcal F_t\right].
$$
With the help of the It\^o lemma (\ref{wzór Ito}) and  because of that the drift of process $F^{T,S}$ equals $0$ under the $T$-\emph{forward martingale measure}, we obtain the diffusion equation for the \emph{forward price}:
$$
\textrm{d}F^{T,S}_t =
\left((\sigma^S_t - \sigma^T_t\right) F^{T,S}_t d{W^T_t}
= -\sigma \left[b(S-t) - b(T-t)\right] F^{T,S}_t d{W^T_t}.
$$
Therefore, the process of \emph{forward prices} is a geometric Brownian motion with $\sigma^{T,S}(t)= -\sigma \left[b(S-t) - b(T-t)\right]$. The random variable $X := \ln \frac{F^{S,T}_T}{F^{S,T}_t}$ has the Gaussian distribution.
Similarly, we obtain the formulas:
\begin{gather*}
X = - \frac12 \int_t^T \sigma^{T,S}(u)^2 du
+ \int_t^T \sigma^{T,S}(u) d{W_u^T},\\
E(X) = -\frac12 \int_t^T \sigma^{T,S}(u)^2 du = -\frac12
\frac{\sigma^2}{\kappa^3} (1 - e^{-\kappa [S - t]})^2
(1 - e^{-2\kappa [T - t]}),\\
v(t,T,S) := \sqrt{\int_t^T \sigma^{T,S}(u)^2 du} =
\frac{\sigma}{\kappa^{3/2}} (1 - e^{-\kappa [S - t]}) \sqrt{1 -
e^{-2\kappa [T - t]}}.
\end{gather*}
Our further reasoning is the same like in the Merton model. Let the function $g: \mathbb{R} \times \mathbb{R_+}
\ni (x,y) \longmapsto \max(e^x y - K,0)$ and the random variable $Z =e^X$. 
We obtain the formula
\begin{align*}
C_t &= B^T_t \left(E^{P^T}\left[\max(F^{T,S}_T
- K,0)|\mathcal F_t\right]\right)(r)\\
&= B^T_t \left(E^{P^T}\left[g(X,F^{T,S}_t)|\mathcal
F_t\right]\right)(r).
\end{align*}
With the help of independence $X$ of $\mathcal F_t$ and $\mathcal F_t$-\emph{measurability} of $F^{T,S}_t$, we can use the lemma (\ref{borelowska}). Then, if we use the lemma (\ref{lognorm}), we calculate that the expected value $E(g(X,y))$ equals 
\begin{multline*}
E(g(X,y)) = y N\left(\frac{1}{v(t,T,S)} \ln\frac{y}{K} + \frac12
v(t,T,S)\right) +\\ - K N\left(\frac{1}{v(t,T,S)} \ln\frac{y}{K} -
\frac12 v(t,T,S)\right).
\end{multline*}
If we substitute to the above equation $y = F^{T,S}_t$ we obtain
\begin{multline*}
\left(E^{P^T}\left[g(X,F^{T,S}_t)|\mathcal F_t\right]\right)(r) =\\
= F^{T,S}_t N\left(\frac{1}{v(t,T,S)} \ln\frac{F^{T,S}_t}{K}
+ \frac12 v(t,T,S)\right) +\\ - K N\left(\frac{1}{v(t,T,S)}
\ln\frac{F^{T,S}_t}{K} - \frac12 v(t,T,S)\right)
\end{multline*}
and finally
\begin{equation*}
C_t = B^S_t N(d_1) - K B^T_t N(d_2),
\end{equation*}
where
\begin{align*}
d_1 &= \frac{1}{v(t,T,S)} \ln\left(\frac{B^S_t}{K B^T_t}\right) + \frac12 v(t,T,S),\\
d_2 &= \frac{1}{v(t,T,S)} \ln\left(\frac{B^S_t}{K
B^T_t}\right)
- \frac12 v(t,T,S),\\
v(t,T,S) &= \frac{\sigma}{\kappa^{3/2}} (1 - e^{-\kappa [S - t]})
\sqrt{1 - e^{-2\kappa [T - t]}}.
\end{align*}

\section{Final remarks}
The Fig.~1 shows the prices of European call options on the zero-coupon bonds with the expiration date $5$ years, the face value $1$ and the real values of the market parameters based on the Merton model and the Vasicek model.
\begin{figure}[h]
\begin{center}
\includegraphics[width=12cm,height=4cm]{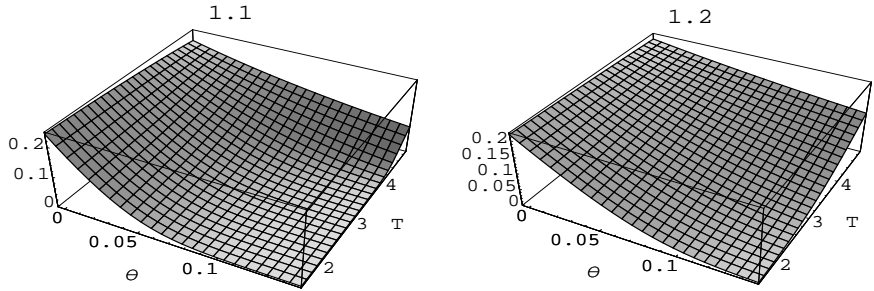}
\end{center}
\caption{The prices of European call options on the zero-coupon bonds based on the: $1.1$-- Merton model and $1.2$--Vasicek model. We assume that: $\kappa = 0.4$, $\sigma = 0.03$, $r=0$.}
\end{figure}
In the above figures, we consider the European call option on the zero-coupon bond with the exercise price $K=0.80$, time to maturity $S=5$ years and the face value $B^S_S=1$. 
As it follows from the general affine model pricing, see (\ref{def:rów stoch}), we can compare this two models under the assumption that $\varphi =\kappa \theta $. The results are qualitatively different. Let us notice that, the price of the option described by Vasicek is higher than the price described by Merton and faster drawn towards to $0.2$, where $0.2$ is the expected price of the option with fixed expiration date $T=5$, because the market is effective without an opportunity of arbitrage. Let us observe that, both models give equal prices for the long interest rate $\theta=0$. \\
 
The difference between the logarithmic price of the European call option described by the Ornstein-Uhlenbeck process and the logarithmic price of the call option described by the
Wiener-Bachelier process is illustrated in the Fig.~2.
\begin{figure}[h]
\begin{center}
\includegraphics[width=8cm,height=6cm]{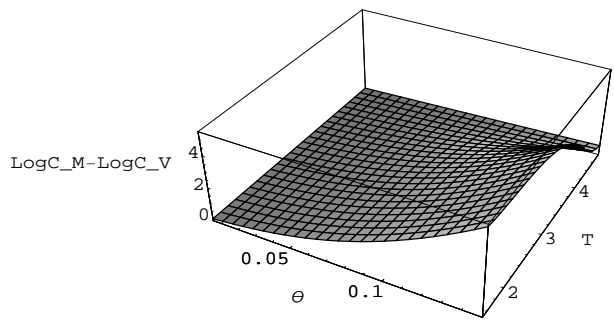}
\end{center}
\caption{We assume that: $\kappa = 0.4$, $\sigma = 0.03$, $r=0$, and $C_M$ is the option price based on the Merton model, and $C_V$ is the option price based on the Vasicek model.}
\end{figure}
The difference between these logarithmic prices increases with the growth of $\theta$ and fades away in the limit of the expiration date ($T=5$) of the zero-coupon bond. One of the inappropriate properties of the Merton model is the constant drift of the short rate. With the constant positive drift the short rate is expected to increase in the future. But this is not realistic. Many empirical studies show that, the interest rates exhibit mean reversion, in the sense that, if an interest rate is high by historical standards, it typically will be falling in the near future (analogically if the current interest rate is low).\\

In paper \cite{7} we considered the prices of options on company's stock supported by Wiener-Bachelier model and Ornstein-Uhlenbeck model and we interpreted the Ornstein-Uhlenbeck process in terms of quantum market games theory as a non-unitary thermal tactic. We compared the probability density functions of both distributions and showed that the
Ornstein-Uhlenbeck process would give the true reflection of the market in time scale where the
approximation by a Wiener-Bachelier process is not valid, but such that market expectations and the asymptotic equilibrium state have not been changed. We called this time interval mezzo-scale. The differences between a classical look and the pricing which is
supported by the quantum model; they are visible for very liquid financial instruments. The quantum game theory takes advantage the Ornstein-Uhlenbeck process for the financial modelling. Quantum strategies create unique opportunities for making profits during intervals shorter than characteristic thresholds for an effective market (Brownian motion). The additional possibilities offered by quantum strategies can lead to more successful outcomes than purely classical ones, see \cite{piot}. 
\newpage
\section{Appendix}

In this section we introduce definitions and theorems (without proofs) which are used in our paper. They can be find in \cite {10},\cite {1},\cite{8},\cite{9},\cite{11}.

\begin{defin}
${}$\\ Let $X$ be an integrable random variable on a probability space $(\Omega,\mathcal F,P)$, and let $\mathcal G$ be a $\sigma$-field contained in $\mathcal F$. Then the {\it conditional expectation value} of $X$, for given $\mathcal G$, is defined to be a random variable $E(X|\mathcal G)$ such that:
\begin{itemize}
\item $E(X|\mathcal G)$ is $\mathcal G$-measurable,
\item for any $A \in \mathcal G$
   $$
   \int_A X dP = \int_A E(X|\mathcal G) dP.
   $$
\end{itemize}
\end{defin}

\begin{lemat}\label{borelowska}
${}$\\If $\mathcal G \subset \mathcal F$ is $\sigma$-field and $f: \mathbb{R} \times \mathbb{R}_{\textbf{+}} \longrightarrow \mathbb{R}$ is a Borel and bounded function, $X$ is a random variable which is independent of $\mathcal G$, and $Y \geq 0$ is $\mathcal G$-measurable and $g(y) = E(f(X,y))$, then
$$
E(f(X,Y)|\mathcal G) = g(Y).
$$
\end{lemat}

\begin{twr}[\textbf{It\^o lemma}]\label{wzór Ito}${}$
\begin{enumerate}[(i)]
\item 
   Let $X=(X_t)_{t \geq 0}$ be a real-valued process with dynamics
   $$
   dX_t = a_tdt + b_t d{W_t},
   $$
   where $a,b$ are real-valued processes, and $W$ is a one-dimensional standard Brownian motion. Then, for any function $F: [0,\infty) \times \mathbb{R} \ni (t,x) \longrightarrow F(t,x) \in \mathbb{R}$ which is two times continuously differentiable in $x$ and continuously differentiable at $t$, the process defined by $Y = (F(t,X_t))_{t \geq 0}$ is an It\^o process with dynamics  
   \begin{multline*}
   dY_t = \left(\frac{\partial F}{\partial t}(t,X_t) + a_t \frac{\partial F}{\partial x}
   (t,X_t) + \frac 12 b_t^2 \frac{\partial^2 F}{\partial x^2}(t,X_t)\right)dt +\\
   + b_t \frac{\partial F}{\partial x}(t,X_t) d{W_t}.
   \end{multline*}
\item 
   Let $W$ be a one-dimensional standard Brownian motion, $a = (\alpha,\beta)$,
   $b = (\gamma,\delta)$ and $(X,Y)$ be two-dimensional stochastic processes with dynamics: 
   \begin{gather*}
   \textrm{d}X_t = \alpha_t dt + \beta_t d{W_t},\\
   \textrm{d}Y_t = \gamma_t dt + \delta_t d{W_t}.
   \end{gather*}
   Then, for any function $F: [0,+\infty) \times \mathbb{R}^{2} \ni (t,x,y) \longrightarrow F(t,x,y)\in \mathbb{R}$ which is two times continuously differentiable in $x$ and $y$ and continuously differentiable at $t$, the process $Z = (F(t,X_t,Y_t))_{t \geq 0}$ is an It\^o process with dynamics
   \begin{multline*}
   \textrm{d}Z_t = \left[\frac{\partial F}{\partial t}(t,X_t,Y_t) + \alpha_t\frac{
   \partial F}{\partial x}(t,X_t,Y_t) + \gamma_t \frac{\partial F}{\partial y}(t,X_t,Y_t)\right. +\\
   + \frac12 \left(\beta_t^2 \frac{\partial^2 F}{\partial x^2}(t,X_t,Y_t)
   + 2 \beta_t \delta_t \frac{\partial^2 F}{\partial x\partial y}(t,X_t,Y_t)\right. +\\
   +\left.\left.\delta_t^2 \frac{\partial^2 F}{\partial y^2}(t,X_t,Y_t)\right)\right]dt
   + \left(\beta_t \frac{\partial F}{\partial x}(t,X_t,Y_t) + \delta_t
   \frac{\partial F}{\partial y}(t,X_t,Y_t)\right)d{W_t}.
   \end{multline*}
\end{enumerate}
\end{twr}

\begin{defin}\label{def:geom Brown}
${}$\\ A stochastic process $X = (X_t)_{t\geq 0}$ is said to be the {\it geometric Brownian motion}\/ if it is a solution of the stochastic differential equation 
$$
\textrm{d}X_t = \mu(t) X_t dt + \sigma(t) X_t d{W_t},
$$
where $\mu,\sigma : [0,+\infty) \longrightarrow \mathbb{R}$ are deterministic functions of time. The initial value for process is assumed to be positive, $X_0 > 0$. 

The function $\mu$ is the growing rate (drift) and $\sigma$ is a volatility rate of the process $X$.
\end{defin}
\begin{twr}\label{twr:geom Brown}
${}$\\Let $X$ be the geometric Brownian motion with $\mu\equiv 0$, then for any $t > 0$ the following conditions are fulfilled:
\begin{enumerate}[(i)]
\item
$X_t > 0$,
\item
a random variable $\ln \frac{X_t}{X_s}$ has a normal distribution,
\item
if $\sigma$ is a volatility rate of the process $X$, then 
\begin{gather*}
X_t = X_s \exp\left((-\frac12 \int_s^t \sigma(u)^2 du
+ \int_s^t \sigma(u) d{W_u}\right),\\
E\left(\ln \frac{X_t}{X_s}\right) = -\frac12 \int_s^t \sigma(u)^2 du,\\
\Var\left(\ln \frac{X_t}{X_s}\right) = \int_s^t \sigma(u)^2 du.
\end{gather*}
\end{enumerate}
\end{twr}

\begin{lemat}\label{lognorm}
${}$\\If $Y = e^X$ and $X \backsim N(m,s^2)$, then for any constant $K > 0$ 
\begin{align*}
E[\max(Y - K,0)] &= e^{m+\frac12 s^2}
N\left(\frac{m - \ln K}{s} + s\right)
- K N\left(\frac{m - \ln K}{s}\right) =\\
&= E(Y) N\left(\frac{m - \ln K}{s} + s\right)
- K N\left(\frac{m - \ln K}{s}\right).
\end{align*}
\end{lemat}

\begin{twr}\label{opcje}
Let $T$ be the expiration date and $K$ the exercise price of the European option, and $h_T$ be the price of the underlying assets at the expiration date, then the price of the option at the time $t < T$ is equal to
\begin{equation}\label{opcje!}
C_t = B^T_t E^{P^T}\left[\max(h_T - K,0)| \mathcal F_t\right].
\end{equation}%
\end{twr}

The basic tool for pricing the financial models is the {\it Feynman-Kac theorem}\/. We assume that $x=(x_t)_{t\geq 0}$ is a diffusion process with dynamics given by the stochastic differential equation 
\begin{equation}
\textrm{d}x_t = \alpha(x_t,t) dt + \beta(x_t,t) d{W^*_t}.
\end{equation}
Moreover, we assume that a process $r$ ($r_t = r(x_t,t)$) describes the short interest rate and the risk-neutral probability measure exists.\par
Let us consider a security with a single payment of $H_T = H(x_T,T)$ and the expiration date $T$. Let $V_t = V(x_t,t)$ denotes the price of the security at the moment $t$. It is proved that 
\begin{equation}
V_t = E^{P^*}\left[e^{-\int_t^T r(x_s,s) ds}
H(x_T,T)|\mathcal F_t\right].
\end{equation}

\begin{twr}[\textbf{Feynman-Kac theorem}]\label{twierdzenie feymana kaca}
${}$\\The function $V$ defined by 
$$
V(x,t) = \left(E^{P^*}\left[e^{-\int_t^T r(x_s,s) ds}
H(x_T,T)|\mathcal F_t\right]\right)(x)
$$
satisfies the partial differential equation 
\begin{equation}
\frac{\partial V}{\partial t}(x,t) + \alpha(x,t) \frac{\partial V}{\partial x}(x,t)
+\frac12 \beta(x,t)^2 \frac{\partial^2 V}{\partial x^2}(x,t) - r(x,t) V(x,t) = 0,
\end{equation}
together with the terminal condition $V(x,T) = H(x,T)$. The process $V_t = V(x_t,t)$
describes the price of the considered security. The process
$(V_t)_{t>0}$ is a diffusion process under $P^*$ and $W^*$ with the drift
\begin{equation}
\mu(x,t) = r(x,t) V(x,t)
\end{equation}
and the volatility rate
\begin{equation}
\sigma(x,t) = \frac{\partial V}{\partial x}(x,t) \beta(x,t).
\end{equation}
\end{twr}

\end{document}